\documentclass[epsf,10pt]{article}
\usepackage{amssymb}
\usepackage{graphicx}
\begin{document}

\makeatletter
\def\@maketitle{\newpage
 \null
 {\normalsize \tt \begin{flushright} 
  \begin{tabular}[t]{l} \@date  
  \end{tabular}
 \end{flushright}}
 \begin{center} 
 \vskip 2em
 {\LARGE \@title \par} \vskip 1.5em {\large \lineskip .5em
 \begin{tabular}[t]{c}\@author 
 \end{tabular}\par} 
 \end{center}
 \par
 \vskip 1.5em} 
\makeatother
\topmargin=-1cm
\oddsidemargin=1.5cm
\evensidemargin=-.0cm
\textwidth=15.5cm
\textheight=22cm
\setlength{\baselineskip}{16pt}
\title{The World-Line Quantum Mechanics Model at Finite Temperature which 
is Dual to the Static Patch Observer in de Sitter Space}
\author{Ryuichi~{\sc Nakayama}
       \\[1cm]
{\small
    Division of Physics, Graduate School of Science,} \\
{\small
           Hokkaido University, Sapporo 060-0810, Japan}
}
\date{
EPHOU-11-009  \\
December  2011  
}
%
%
\maketitle

\begin{abstract} 
A simple conformal quantum mechanics model of a d-component variable is 
proposed, which exactly reproduces the retarded Green functions and conformal 
weights of conformally coupled scalar fields in de Sitter spacetime seen 
by a static patch observer. It is found that the action integral of this 
model is automatically expressed by a complex integral over the time variable 
$t$ along a closed contour in a way which is typical to the Schwinger-Keldysh 
formalism of a thermofield theory.  Hence this model is at finite 
temperature. The case of conformally coupled scalar 
fields in 3d Schwarzschild de Sitter space is also considered and then a 
large-N matrix model is obtained.

\end{abstract}
\newpage
\setlength{\baselineskip}{18pt}

\newcommand {\beq}{\begin{equation}}

\newcommand {\eeq}{\end{equation}}
\newcommand {\beqa}{\begin{eqnarray}}
\newcommand {\eeqa} {\end{eqnarray}}
\newcommand{\bm}[1]{\mbox{\boldmath $#1$}}
\newcommand{\Sq}{D(X)}
\newcommand{\al}{2\pi \alpha'}
\newcommand{\RR}{{\mathsf R\hspace*{-0.9ex}%
  \rule{0.15ex}{1.5ex}\hspace*{0.9ex}}}
\section{Introduction}
\hspace{5mm}
Compared to the well-established AdS/CFT correspondence,\cite{Mardacena}
\cite{Gubser}\cite{Witten} the de Sitter 
holography is less understood. At least there are two kinds of de Sitter 
duality which depend on two types of observers. In the global patch 
of de Sitter spacetime the boundaries are at the future and past 
infinity ${\cal I}^+$, ${\cal I}^-$, and the bulk gravity theory is 
dual to the conformal field theory at ${\cal I}^+$. The metaobserver 
at ${\cal I}^+$ can see all the events there through the wave function of 
the universe\cite{HH}. This type of holography has been developed by means of 
analogy to AdS/CFT duality. \cite{Witten1}\cite{Strominger}\cite{Mardacena2}
\cite{ANS}\cite{AHS}

In the static patch the observer is surrounded by a cosmological horizon.
The future infinity ${\cal I}^+$ is behind the cosmological horizon and 
one cannot use ${\cal I}^+$ for the holography.\cite{GKS}\cite{Banks}
\cite{PV}\cite{BFM}\cite{Suss}\cite{CLM} In this case the dual 
description of the de Sitter spacetime may be expected for the static 
patch observer residing at $r=0$. Then, because the (d-1)-sphere in the 
static patch collapses at $r=0$, dS space may be described by a conformal 
worldline quantum mechanics at the center of dS space.\cite{AHH} 

In \cite{AHH} the retarded Green function of scalar field and 
gravitational fluctuations seen by a static patch observer are studied. 
For conformally coupled scalar fields and gravitons in four dimensional 
space it was shown that the retarded Green functions take the scaling 
form and the scaling exponents and wave functions are controlled by a 
hidden $SL(2,{\mathbb R})$ symmetry. In the general case without conformal 
invariance it was also observed that there are underlying $SL(2,{\mathbb R}) 
\times SL(2,{\mathbb R})$ symmetry. They also show how the worldline de 
Sitter propagators can be reproduced from conformal quantum mechanics and 
speculate that the static patch of de Sitter spacetime may be dually 
described in terms of large N quantum mechanics. They also presented a 
simple free model of large N quantum matrix model. The conformal weights 
derived from this matrix model in the diagonal approximation, however, do 
not exactly agree with the de Sitter results for scalar fields with 
non-conformal coupling $x \neq \frac{1}{2}$.

In this paper a simple conformal quantum mechanics model of d-component 
variables, $x_i(\tau)$ ($i=1,..,d$), is proposed, which exactly reproduces 
the retarded Green functions and conformal weights of scalar fields 
with conformal coupling and four dimensional gravitons. First, the model of 
non-matrix type will be obtained, and then a generalization to 
a large N matrix model will be also presented. The case of scalar fields in 
3d Schwarzschild-de-Sitter space with conformal coupling is also considered.

In de Sitter space an observer is in a thermal bath of particles and the 
de Sitter space is associated with a de Sitter temperature. Thus the conformal 
quantum mechanics constructed in this paper is expected to be associated with 
this temperature. It will be shown that this is indeed the case by 
demonstrating that the action integral for this quantum mechanics is 
represented as a contour integral along a contour in the complex-time plane 
as in the thermofield theory. 

This paper is organized as follows. In sec.2 the scaling form of retarded 
Green function in static patch de Sitter space is reviewed. In sec.3
the conformal quantum mechanics is briefly reviewed. In sec.4 a new 
conformal quantum mechanics theory which reproduces the conformal weights of
the retarded Green function in de Sitter space seen by a static patch oberver 
is presented. Then the primary states with eigenvalues for the conformal 
generator $R$ is presented. Lagrangian and Hamiltonian of this conformal 
quantum mechanics reexpressed in terms of de Sitter time $t$ for the static 
patch are presented in sec.5.  Interestingly, it is shown that the action 
integral is expressed as a contour integral along a closed contour $C$ in 
the complex-$t$ plane. Hence the dynamical variables also live on a line 
parallel to the real axis. This is reminiscent of the Schwinger-Keldysh 
formalism of thermofield theory. The de Sitter temperature is 
naturally encoded in the conformal quantum mechanics model. 
In sec.6 3d Schwarzschild-de Sitter space is considered and the conformal 
quantum mechnics model which reproduce the conformal weight is considered.  
In sec.7 a large N matrix model which is dual 
to the static patch observer is presented. Sec.8 is left for discussions.

\section{Retarded Green functions and Conformal Weights}
\hspace{5mm}
In \cite{AHH} the propagators of scalar fields seen by a static patch 
observer in de Sitter spacetime were studied. The static patch coordinate 
system for (d+1)-dimensional de Sitter spacetime is given by 
\begin{eqnarray}
ds^2=-f(r) dt^2+\frac{1}{f(r)} \, dr^2+r^2 \, d\Omega^2_{d-1},
\end{eqnarray} 
where 
\begin{eqnarray}
f(r)= 1-\left(\frac{r}{\ell}\right)^2
\end{eqnarray}
with $\ell$ related to the cosmological constant $\Lambda$ by 
$\Lambda=(d-1)(d-2)/2\ell^2$ and $d\Omega_{d-1}^2$ the round metric on 
$S^{d-1}$. There is a cosmogical horizon at $r=\ell$ and the observer is 
at $r=0$. The symmetry of this metric is $SO(d) \times {\mathbb R}_t$.

In \cite{AHH} the wave equation for a free scalar field in the static patch, 
$\Delta \Phi =m^2 \, \Phi$, was studied and its retarded Green function was 
obtained. The mass $m$ of the scalar field is parametrized by 
\begin{eqnarray}
\ell^2 \, m^2=\frac{d^2}{4}-x^2 \label{x}
\end{eqnarray}
and $x=1/2$ corresponds to the conformal coupling.  
In the case of conformally coupled scalar fields the retarded Green function 
of the operator of an $SO(d)$ angular momentum $l$ was shown to have the form.
\begin{eqnarray}
G^R_l(t) \propto \theta(t) \left( \frac{\ell^{-1}}
{\sinh \frac{1}{2}t/\ell}\right)^{2\Delta}, 
\label{retarded}
\end{eqnarray}
where the conformal weight was found to be
\begin{eqnarray}
\Delta= l+\frac{1}{2} \, (d-1).
\label{sdimension}
\end{eqnarray}

The poles of the retarded Green functions with a complex frequency $\omega$ 
give the quasi-normal modes. It was shown that the wave functions of the 
quasi-normal modes are controlled by a hidden $SL(2,{\mathbb R})$ symmetry. The 
conformal weight (\ref{sdimension}) is found to be related to the Casimir 
operator ${\cal C}$ of the $SL(2,{\mathbb R})$ algebra acting on the wave 
functions of 
the quasi-normal modes by the formula ${\cal C}=\Delta (\Delta-1)$. This 
allows one to associate each mode $l$ of the scalar field 
to a dual operator with conformal weight $\Delta$.  A similar analysis for 
the gravitational perturbation was also reported. Especially in four 
dimensional de Sitter spacetime the scaling dimesions of the operators are 
given by exactly the same expression as (\ref{sdimension}). 

It is also pointed out that this $SL(2,{\mathbb R})$ symmetry has an origin in 
$AdS_2 \times S^{d-1}$, which is the conformally rescaled metric of de Sitter 
spacetime. 

\section{Conformally Invariant Quantum Mechanics}
\hspace{5mm}
In \cite{AHH} it was also studied how the structure of the retarded Green 
function in the case of the conformally coupled scalar fields is reproduced 
by a dual conformal quantum mechanics theory on the worldline on the 
static observer in de Sitter spacetime. A free theory example of a 
conformal quantum mechanics theory of a large N matrix was prorposed. 
By analogy with the established AdS/CFT dualities it is 
expected in \cite{AHH} that the dual worldline theory should be a large N 
matrix theory. It was shown that in this free theory the scaling 
dimension, or the lowest eigenvalues of an operator $R$, which will be 
explained below, (\ref{RLL}), takes the value $r_0=dN/2+l/2$. Here $N$ is 
the number of eigenvalues of the matrix variable.  It is argued that this 
value $r_0$ corresponds to a lowest weight representation of 
$SL(2,{\mathbb R}) \times SL(2,{\mathbb R})$. However, the exact coincidence
 was not observed. 

In the following section a conformal quantum mechanics theory of a 
$d$-component 
variable $x_i$, ($i=1,2,...,,d$) will be presented and it will be shown 
that this theory reproduces the conformal weight (\ref{sdimension}). 
This theory is not a large N-matrix theory. However, an extension to a 
matrix theory will be discussed in sec 6. Meanwhile, in this section 
conformally invariant quantum mechanics theory will be briefly reviewed. 

Conformally invariant quantum mechanics was studied in \cite{DAFF}, and 
recently in \cite{Jackiw} in the context of $AdS_2/CFT_1$ correspondence. 
The simplest model is the one with the coordinate $q(\tau)$. $\tau$ is the 
time variable. The Lagrangian is given by 
\begin{eqnarray}
L= \frac{1}{2} \, \dot{q}^2-\frac{g}{2q^2}.
\end{eqnarray}
Here $\dot{q}=dq/d\tau$ and $g$ is a dimensionless constant. This model is 
invariant under a time translation ($H: \tau \rightarrow \tau +\epsilon$), 
dilatation ($D: \tau \rightarrow \tau +\epsilon \tau$) and conformal 
transformation ($K: \tau \rightarrow \tau+ \epsilon \tau^2$). ($\epsilon$ is an 
infinitesimal constant parameter.) Under these transformations $q(\tau)$ 
changes as follows.
\begin{eqnarray}
\delta_H q &=& \epsilon \, \partial_{\tau} q, \\
\delta_D q &=& \epsilon (\tau \partial_{\tau}q-\frac{1}{2} \, q), \\
\delta_K q &=& \epsilon (\tau^2 \, \partial_{\tau} q-\tau q)
\end{eqnarray}
The action integral is invariant, and the corresponding conserved charges 
$H$, $D$ and $K$ satisfy the $SL(2,{\mathbb R})$ algebra. 
\begin{eqnarray}
&& [D, K]= iK, \nonumber \\
&& [H, K]= 2iD, \nonumber \\
&& [H,D] = iH \label{SL2R}
\end{eqnarray} 

The charges can be presented in another basis.
\begin{eqnarray}
R = \frac{1}{2} \, (\ell H+\ell^{-1}K), \qquad L_{\pm}= \pm i D+\frac{1}{2} (\ell^{-1}K-\ell H)
\label{RLL}
\end{eqnarray}
The new charges satisfy 
\begin{eqnarray}
[R, L_{\pm}]=\pm L_{\pm}, \qquad [L_-,L_+]=2R. 
\end{eqnarray}

The operator $R$ can be taken to be a positive operator and the above algebra 
shows that its eigenvalues are integrally spaced. So $R$ can be used to 
classify the normalizable states. $L_+$ and $L_-$ are raising and lowering 
operators. The eigenstates of $R$ are defined by 
\begin{eqnarray}
R \, |r \rangle=r \, |r \rangle, 
\end{eqnarray}
and the primary states satisfy an additional constraint. 
\begin{eqnarray}
 L_- \, |r_0  \rangle=0
\end{eqnarray}
The eigrnvalue of $R$, $r$, takes the values $r=r_n=r_0+n$, ($n=0,1,2, ...$). 

The Lie algebra (\ref{SL2R}) posseses a Casimir operator.
\begin{eqnarray}
{\cal C}= \frac{1}{2} \, (HK+KH)-D^2
\label{Casimir}
\end{eqnarray}
On the tower of eigenstates $|r_n \rangle$ it takes a constant value.
\begin{eqnarray}
{\cal C} \, |r_n \rangle = r_0 \, (r_0-1) \, |r_n \rangle 
\label{Casimirr0}
\end{eqnarray}

The $SL(2,{\mathbb R})$-invariant time evolution of the primary state $|r_0 
\rangle$ with respect to $H$ can be introduced as follows.
\begin{eqnarray}
|r_0, \tau \rangle= N(\tau) \, e^{-\omega(\tau) \, L_+} \, |r_0 \rangle
\label{stater0}
\end{eqnarray}
Here the functions are
\begin{eqnarray}
N(\tau) = \sqrt{\Gamma(2r_0)} \, \left(\frac{\omega(\tau)+1}{2} 
\right)^{2r_0}, \qquad \omega(\tau)=\frac{\ell+i\tau}{\ell-i\tau}.
\end{eqnarray}
One can show that this time-evolved state starts at an imaginary time. 
\footnote{To prove this one should use (3.17) and (3.15) of \cite{Jackiw} to 
show $\exp (-\ell H) \exp (-L_+) |r_0 \rangle \propto |r_0 \rangle$. 
(Notations in \cite{Jackiw} are different from those here.) This means 
$\exp (-L_+) |r_0 \rangle \propto \exp (\ell H) |r_0 \rangle$. Then by using 
$|r_0, \tau \rangle= \exp (iH\tau) |r_0, 0\rangle$ and $|r_0, 0 \rangle \propto 
\exp (-L_+)|r_0\rangle$, the latter of which derives from (\ref{stater0}), one 
obtains (\ref{imaginary time}).} 
\begin{eqnarray}
|r_0, \tau \rangle \propto e^{iH\tau} \, e^{\ell H} \, |r_0\rangle
\label{imaginary time} 
\end{eqnarray}
By expanding the state (\ref{stater0}) in terms of $|r_n \rangle$ the 
correlator is calculated as
\begin{eqnarray}
\langle r_0, \tau' |r_0, \tau \rangle= 
\frac{\Gamma(2r_0) \, \ell^{2r_0}}{[2i(\tau'-\tau)]^{2r_0}}. 
\label{correlator1}
\end{eqnarray}

This does not agree yet with the scaling correlator obtained from the retarded 
Green function (\ref{retarded}) of the conformally coupled scalar field in 
de Sitter space. In addition to the \lq scaling time' $\tau$, 
it is necessary to introduce \lq de Sitter time' 
$t$ defined by \cite{AHH} 
\begin{eqnarray}
\tau =\ell \, \tanh \frac{t}{2\ell}.
\label{deSittertime}
\end{eqnarray}
The evolution along time $t$ is generated by a new Hamiltonian \cite{AHH}
\begin{eqnarray}
H_0=\frac{1}{2} \, ( H-\ell^{-2} K)=H- \ell^{-1}R. \label{H0}
\end{eqnarray}
 The relation 
(\ref{deSittertime}) is substituted into (\ref{correlator1}). 
Then one obtains 
\begin{eqnarray}
\langle r_0, \tau' |r_0, \tau \rangle= \frac{\Gamma(2r_0)}{(2i)^{2r_0}} \ 
\left[ \frac{\cosh \frac{t'}{2\ell} \, \cosh \frac{t}{2\ell}}
{\sinh \frac{1}{2\ell} \, (t-t')} \right]^{2r_0}.
\end{eqnarray}
This is still different from (\ref{retarded}).
For an exact agreement the state $|r_0, \tau\rangle$ must be rescaled 
by a certain factor, 
\begin{eqnarray}
|r_0,t \rangle_{dS}= (2d\tau/dt)^{r_0} \, |r_0,\tau \rangle=
\left[1-\left(\frac{\tau}{\ell}\right)^2\right]^{r_0} \, |r_0,\tau \rangle .
\label{rescale}
\end{eqnarray}
Then one obtains
\begin{eqnarray}
{}_{dS} \langle r_0, t'|r_0,t \rangle_{dS}= \frac{\Gamma(2r_0)}{(2i)^{2r_0}} \ 
\left[ \frac{1}{\sinh \frac{1}{2\ell} \, (t-t')} \right]^{2r_0} .
\label{correlator2}
\end{eqnarray}
This coincides with (\ref{retarded}) and (\ref{sdimension}), if one sets 
$r_0=\Delta$. 

\section{Quantum Mechanics Model Dual to the Static Patch Observer}
\hspace{5mm}
In this section we will construct a conformal quantum mechanics model which 
reproduces the $R$ eigenvalues of the primary state, $r_0=\Delta=l+(d-1)/2$. 

Let us consider a model, which is described by d-vector $\bm{x}$ with 
components, $x_i(\tau)$ $(i=1,..,d)$. The Lagrangian is given by 
\begin{eqnarray}
L= \frac{1}{2} \, \dot{\bm{x}}^2  -\frac{g_0}{2} \, 
\left[ \dot{\bm{x}}^2-\frac{(\bm{x} \cdot \dot{\bm{x}})^2}{\bm{x}^2}
\right] -\frac{g_1}{\bm{x}^2}. \label{29}
\end{eqnarray} 
Here $g_0$, $g_1$ are dimensionless constants. Note also 
$\dot{x}_i= d x_i/d\tau$. The momentum conjugate to $x_i$ is given by 
\begin{eqnarray}
p_i=  \dot{x}_i-g_0\left[ \dot{x}_i
-\frac{\bm{x} \cdot \dot{\bm{x}}}{\bm{x}^2} \, x_i \right].
\end{eqnarray}
This relation can be converted into the following relations.
\begin{eqnarray}
&&\dot{\bm{x}}^2 = \frac{1}{(1-g_0)^2} \, \left[ \bm{p}^2
+\frac{g_0(g_0-2)}{\bm{x}^2} \, (\bm{x} \cdot \bm{p})^2\right], \\
&&\frac{1}{ \bm{x}^2} \, (\bm{x} \cdot \dot{\bm{x}})^2 
= \frac{1}{\bm{x}^2} \, (\bm{x} \cdot \bm{p})^2
\end{eqnarray}
By using these equations one can compute the Hamiltonian.
\begin{eqnarray}
 H=\frac{1}{2(1-g_0)} \, \bm{p}^2-\frac{g_0 }{2(1-g_0)} \, \frac{(\bm{x} 
\cdot \bm{p})^2}{\bm{x}^2}+\frac{g_1}{\bm{x}^2} \label{H}
\end{eqnarray}

This theory has an $SL(2,{\mathbb R})$ symmetry. Under $SL(2,{\mathbb R})$ 
transformations $x_i(\tau)$ changes as follows.
\begin{eqnarray}
\delta_H x_i &=& \epsilon \, \partial_{\tau} x_i, \nonumber  \\
\delta_D x_i &=& \epsilon (\tau \partial_{\tau}x_i-\frac{1}{2} \, x_i), 
\label{HDK} \\
\delta_K x_i &=& \epsilon (\tau^2 \, \partial_{\tau} x_i-\tau x_i) \nonumber 
\end{eqnarray}
The corresponding conserved Noether charges are obtained by the usual method. 
The Hamiltonian $H$ is given in (\ref{H}). 
\begin{eqnarray}
D &=& \tau \, H-\frac{1}{4} \, (\bm{p} \cdot \bm{x}+\bm{x} \cdot \bm{p}),
 \label{D} \\
K &=& \tau^2 \, H-\frac{1}{2} \, \tau \, (\bm{p} \cdot \bm{x}
+\bm{x} \cdot \bm{p})+\frac{1}{2} \,\bm{x}^2 \label{K}
\end{eqnarray}

Now these charges at $\tau=0$ will be considered. The radial coordinate 
$\rho= \sqrt{\bm{x}^2}$ is introduced. 
Then the above charges are represented as 
\begin{eqnarray}
 H &=& -\frac{1}{2}\, \left[ \frac{\partial^2}
{\partial \rho^2}
+\frac{d-1}{\rho} \frac{\partial}{\partial \rho} \right] 
+ \frac{ l(l+d-2)}{2(1-g_0)\rho^2}+\frac{g_1}{\rho^2}, 
\label{Hr}\\
D &=& i \, \left(\frac{1}{2} \, \rho \frac{\partial}{\partial \rho}
+\frac{d}{4} \right), \label{Dr} \\
K &=& \frac{1}{2} \ \rho^2 \label{Kr}
\end{eqnarray}
Here $l$ is the magnitude of the $SO(d)$ angular momentum and a sector with 
a fixed $l$ is considered. 
The operator ordering in the second term of (\ref{H}) is here defined as 
$(\bm{x}^2)^{-1} \, (\bm{x} \cdot \bm{p})^2=-(\partial_{\rho}^2+(d-1)
\rho^{-1} \partial_{\rho})$. With this definition the 
resulting charges are hermitian and obey the correct $SL(2,{\mathbb R})$ 
Lie algebra (\ref{SL2R}). 

The Casimir operator (\ref{Casimir}) takes a constant value in the sector 
with a fixed $l$. By using (\ref{Hr})-(\ref{Kr}) this 
value is evaluated as
\begin{eqnarray}
{\cal C}= \frac{1}{16} \, d(d-4)+\frac{1}{4} \, \frac{l(l+d-2)}{1-g_0}
+\frac{1}{2} \, g_1.
\end{eqnarray}
The value of ${\cal C}$ is related as in (\ref{Casimirr0}) to the lowest 
eigenvalue of $R$ by ${\cal C}=r_0(r_0-1)$. By considering 
${\cal C}+\frac{1}{4}$, the following equation is obtained
\begin{eqnarray}
(r_0-\frac{1}{2})^2= \frac{1}{16} \, (d-2)^2+\frac{1}{4} \, 
\frac{l(l+d-2)}{1-g_0}+\frac{1}{2} \, g_1.
\end{eqnarray}
Then it is easy to see that in order to have a solution\footnote{In this paper 
the prescription of \cite{AHH} to normalize the correlation functions by the 
one with the lowest scaling weight is not adopted.}
\begin{eqnarray}
r_0=l+\frac{d-1}{2}
\label{r0}
\end{eqnarray}
$g_0$ and $g_1$ must take the following values.
\begin{eqnarray}
g_0 = \frac{3}{4}, \qquad 
g_1 = \frac{3}{8} \, (d-2)^2  \label{g0g1}
\end{eqnarray}

To summarize the model Lagrangian which gives the appropriate lowest weight 
of $R$ is the following.
\begin{eqnarray}
L= \frac{1}{8} \, \dot{\bm{x}}^2+\frac{3}{8} \, \frac{(\bm{x} \cdot 
\dot{\bm{x}})^2}
{\bm{x}^2}-\frac{3}{8} \, \frac{(d-2)^2}{\bm{x}^2}
\label{Lagrangian1}
\end{eqnarray}

The correlater (\ref{correlator1}) of the R-primary states can be 
calculated from (\ref{Lagrangian1}). With the change of time variable to 
the de Sitter time $t$ (\ref{deSittertime}) and rescaling of the state 
(\ref{rescale}), one obtains the correlator (\ref{correlator2}). 

Now the wave functions for the primary states will be computed. 
In the $\rho(=\sqrt{\bm{x}^2})$ representation the charges $R$, $L_{\pm}$ are 
represented as follows.
\begin{eqnarray}
R &=& \frac{1}{2} \, (\ell^{-1} K+ \ell H) \nonumber \\
&=&\frac{1}{4\ell} \, \rho^2-\frac{1}{4} \, \left[ \partial_{\rho}^2
+\frac{d-1}{\rho} \, \partial_{\rho} \right] + \frac{ l(l+d-2)}{\rho^2}
+ \frac{3}{16 \rho^2} \, (d-2)^2, \\
L_{\pm} &=& \pm iD-R+\ell^{-1}K = \mp \frac{1}{2}\left(\rho \, \partial_{\rho} 
+\frac{d}{2}\right)-r_n+\frac{1}{2\ell} \, \rho^2
\end{eqnarray}
The operator $R$ in the second equation is replaced by its eigenvalue $r_n$ 
on $|r_n \rangle$.
Separating variables using $SO(d)$ spherical harmonics $Y_l(\Omega)$, 
\begin{eqnarray}
\Psi(\rho, \Omega)= \psi_{l}(\rho) \, Y_l(\Omega),
\end{eqnarray}
the normalized primary state $\psi_{l0}(\rho)$ which obeys $L_- \psi_{l0}=0$ 
can be solved and given by 
\begin{eqnarray}
\psi_{l0}(\rho)=\sqrt{\frac{\Gamma(d/2)}{\Gamma(2r_0) \, \pi^{d/2}}}
\ \rho^{-\frac{d}{2}} \, \left(\frac{\rho^2}{\ell}\right)^{r_0} \, 
e^{-\frac{\rho^2}{2\ell}}.
\label{primarystate}
\end{eqnarray}
Here $r_0$ is given by (\ref{r0}).
The descendants $\psi_{ln}(\rho)$ ($n \geq 1$) are obtained by applying the 
raising operator $L_+$ repeatedly on $\psi_{l0}(\rho)$, or by solving the 
equation $R \, \psi_{l n}=r_n \, \psi_{ln}$, and are given by the 
associated Laguerre polynomial as in \cite{Jackiw}. 
\begin{eqnarray}
\psi_{ln}(\rho)=\sqrt{\frac{n! \  \Gamma(d/2)}{\Gamma(n+2r_0)
\, \pi^{d/2}}}
\ \rho^{-\frac{d}{2}} \, \left(\frac{\rho^2}{\ell}\right)^{r_0} \, 
e^{-\frac{\rho^2}{2\ell}} \, L^{(2r_0-1)}_n \left(\frac{\rho^2}{\ell}\right)
\end{eqnarray}
These states are the dual to the quasi-normal modes of the scalar fields in 
the static patch of de Sitter spacetime. 
The $R$-eigenvalue of the state (\ref{primarystate}) can be also obtained by 
computing $R \, \psi_{l0}(\rho)$ and agrees with $r_0=l+\frac{d-1}{2}$. 

\section{Transformation to the De Sitter Time $t$ 
and Schiwinger-Keldysh Formalism}
\hspace{5mm}
Up to now the Lagrangian (\ref{Lagrangian1}) of the conformal quantum 
mechanics was expressed in terms of the scaling time $\tau$. 
Now the Lagrangian expressed by using the de Sitter time
 $t$ will be studied. This is done by using the relation (\ref{deSittertime}). 

One must notice that due to the Jacobian factor $d\tau/dt$ the new Lagrangian 
$L_{dS}$ is related to the old one by $L_{dS}=L/ (2 \cosh^2 \frac{t}
{2\ell})$. The variable $x_i(\tau)$ is also rescaled as follows.
\begin{eqnarray}
x_i(\tau)=\left(2 \, \frac{d\tau}{dt}\right)^{\frac{1}{2}} \, \tilde{x}_i(t)=
\left(\cosh^2 \frac{t}{2\ell}\right)^{-\frac{1}{2}} \, \tilde{x}_i(t)
\label{confx}
\end{eqnarray}
After a bit of calculation the new Lagrangian is obtained.
\begin{eqnarray}
L_{dS} &=& \frac{1}{4} \, \left(\frac{d\bm{\tilde{x}}}{dt}\right)^2
+\frac{3}{4} \, \frac{1}{\bm{\tilde{x}}^2} \, \left(\bm{\tilde{x}} \cdot 
\frac{d\bm{\tilde{x}}}{dt}\right)^2
-\frac{3}{16} \, \frac{(d-2)^2}{\bm{\tilde{x}}^2}
+\frac{1}{4 \, \ell^2} \, \tilde{\bm{x}}^2      \nonumber \\
&& -\frac{1}{2\ell} \,\frac{d}{dt} 
\left( \tanh \frac{t}{2\ell} \ \tilde{\bm{x}}^2  \right)
\label{tildeL0}
\end{eqnarray}
The last term can be dropped, being a total derivative. 
By the conformal transformation (\ref{confx}) the kinetic term for $\bm{x}$ 
does not take a $t$-dependent factor. 
The action integral obtained from this Lagrangian should be certainly 
$SL(2,{\mathbb R})$ invariant, because this was derived from the 
invariant Lagrangian (\ref{Lagrangian1}) by a change of the time variable, 
but this symmetry is implicitly realized compared to (\ref{HDK}).
\begin{eqnarray}
\delta_H \, \tilde{x}_i &=& \epsilon \, \left(2\partial_t \tilde{x}_i \, 
\cosh^2 \frac{t}{2\ell} \  
 -\frac{1}{2\ell} \, \tilde{x}_i \, \sinh \frac{t}{\ell} 
\  \right), \nonumber \\
\delta_D \, \tilde{x}_i &=& \epsilon \, \left(  \partial_t \, \tilde{x}_i \, 
 \sinh \frac{t}{\ell} 
-\frac{1}{2\ell} \, \tilde{x}_i \, \cosh \frac{t}{\ell} 
  \right), \label{SL2} \\
\delta_K \, \tilde{x}_i &=& \epsilon \, \left( 2 \, \partial_t \, 
\tilde{x}_i\, \sinh^2 
\frac{t}{2\ell} \ -\frac{1}{2\ell} \ \tilde{x}_i\, 
\sinh \frac{t}{\ell}  \right)   \nonumber 
\end{eqnarray}
Note that the time translation $t \rightarrow t+\epsilon$ corresponds to 
the combination, $(1/2)(\delta_H- \delta_K)$. 
 
The Hamiltonian is found to have the following form. 
\begin{eqnarray}
H_{dS} =  \tilde{\bm{p}}^2-\frac{3}{4}  \, 
\frac{1}{\tilde{\bm{x}}^2}
\, (\tilde{\bm{x}} \cdot \tilde{\bm{p}})^2+ \frac{3}{16} \, (d-2)^2 \, 
\frac{1}{\tilde{\bm{x}}^2}-\frac{1}{4\ell^2} \, \tilde{\bm{x}}^2
=\frac{1}{2} \, \tilde{H}-\frac{1}{2\ell^2} \, \tilde{K} \label{H0tilde}
\end{eqnarray}
$\tilde{\bm{p}}$ is the momentum conjugate to $\tilde{\bm{x}}$. 
$\tilde{H}$ is  (\ref{H}) with (\ref{g0g1}) substituted and $\bm{x}$, 
$\bm{p}$ replaced by $\tilde{\bm{x}}$ and $\tilde{\bm{p}}$. $\tilde{K}$ 
is obtained from (\ref{K}) similarly. With the similarly defined $\tilde{D}$,  
the operators $\tilde{H}$ and $\tilde{K}$ obey the algebra (\ref{SL2R}). 
This hamiltonian (\ref{H0tilde}) has a form equal to (\ref{H0}). 

If the duality is correct, the conformal quantum mechanics must be at 
de Sitter temperature. This is encoded in the 
transformation of the time variables, (\ref{deSittertime}). 
When the de Sitter time is Wick rotated, $t \rightarrow -it_E$, the scaling 
time is given by 
\begin{eqnarray}
\tau = -i \, \ell \, \tan \frac{t_E}{2\ell}. 
\end{eqnarray}
Hence the quantum mechanics theory become periodic in $t_E$ with a period 
$\beta=2\pi \ell$. This is consistent with the expectation that this 
quantum mechanics is at de Sitter temperature $T_{dS}=\beta^{-1}=1/(2\pi \ell)$. 

There is one peculiarity about the time integration contour for the action in 
the model. The inverse relation of (\ref{deSittertime}), 
\begin{eqnarray}
t=\ell \, \log \frac{\ell+\tau}{\ell-\tau}, 
\end{eqnarray}
shows that only the segment of the scaling time, $
-\ell \leq \tau \leq \ell$,  corresponds to the de Sitter time axis  
$-\infty < t < \infty$. The remaining regions of $\tau$, $\tau < -\ell$ 
and $\tau > \ell$, are mapped onto a line $\{t= s- \pi \ell i \, | \, 
-\infty < s < \infty\}$, which is parallel to the 
real axis, on the complex-$t$ plane. Hence the line from $\tau=-\infty$ 
to $\tau=+\infty$ is mapped onto a contour $C$ which goes from $t=-\infty$ to 
$t=-\infty-2\pi \ell i$ with a detour. It is composed of four parts. See Fig.1.
\begin{description}
\item [I.]  a path from $t=-\infty$ to 
$t=+\infty$ along the real axis
\item [II.] a path from $t=+\infty$ to 
$t=+\infty-\pi \ell i$ along a segment 
parallell to the imaginary axis
\item [III.] a path from $t=+\infty -\pi 
\ell i$ to $t=-\infty -\pi \ell i$
\item [IV.] a path from $t=-\infty -\pi 
\ell i$ to $t=-\infty-2\pi \ell i$ 
along a segment  parallel to the imaginary axis. 
\end{description}
This contour $C$ can be made closed by imposing a periodic boundary 
condition $\tilde{\bm{x}}(t+i \, \beta)=\tilde{\bm{x}}(t)$ on the variable 
$\tilde{\bm{x}}$. 



\begin{figure}[htbp]
 \begin{center}
  \includegraphics[width=100mm]{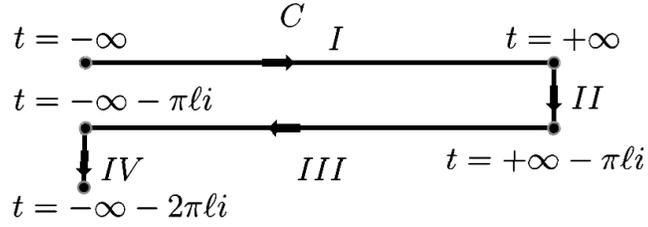}
 \end{center}
 \caption{The contour $C$ for the action 
integral}
 \label{fig:one}
\end{figure}

\vspace*{.5cm}

This contour is reminiscent of the Schwinger-Keldysh formalism of 
thermofield theory.\cite{SK}\cite{HS} The action integral $S$ obtained by 
means of the change of variable (\ref{deSittertime}) from 
$\int_{-\infty}^{\infty}d\tau L(\tau)$ is given by 
\begin{eqnarray}
S = \int_C dt L_{dS}(t)= \int_{-\infty}^{\infty}dt \, L_{dS}(t)
-\int_{-\infty}^{\infty}dt
\, L_{dS}(t-\pi \ell i). \label{Sc}
\end{eqnarray}
The contributions from the segments II, IV are actually absent, being placed at 
infinities.  Hence this conformal quantum mechanics theory turns out to 
contain twice the real degrees of freedom. The first term on the right side of 
the above equation corresponds to the real system and the second to its 
fictitious copy. Actually, the first term will describe the southern causal 
diamond in the global patch \cite{SSV}, where the observer lives, and the 
second term the northern causal diamond beyond the cosmological horizon. 

In a field theory at finite temperature the original system is 
doubled.\cite{SK}\cite{GKS}\cite{HS} The doubled system will be described by 
an entangled state called a thermofield double.
\begin{eqnarray}
|\psi \rangle = \frac{1}{\sqrt{Z}} \, \sum_i \, e^{-\frac{1}{2} \beta E_i} \, 
|E_i\rangle \otimes |E_i\rangle. \label{thermodouble}
\end{eqnarray}
Here $Z(\beta)$ is the partition function. The hamiltonian of the thermofield 
double will be given by $H_{thermo \ double}=H_{dS} 
\otimes I-I \otimes H_{dS}$ and $E_i$ is the eigenvalue of $H_{dS}$. The state
$|\psi \rangle$ is an eigenstate of $H_{thermo \ double}$ with an eigenvalue 
zero. The density matrix $\hat{\rho}$ of the original system is given by 
computing the trace of $|\psi\rangle \langle \psi|$ over the second Hilbert 
space of the tensor product. The Green functions of the theory (\ref{Sc}) 
ordered along the contour $C$, the Schwinger-Keldysh correlators, are known 
to be related to the retarded Green function 
$\tilde{G}^R(\omega)$.\cite{SK}\cite{HS}  

There is, however, a problem. The eigenvalues of $H_{dS}$ are not descrete. 
Furthermore, the potential in (\ref{H0tilde}) is unbounded from below. Then 
the state $|\psi \rangle$ becomes a pure state with $E_i=-\infty$.  
Because the $SL(2, {\mathbb R})$ Casimir is fixed in this theory, this 
problem might be solved by constructing the thermofield 
double $|\psi \rangle$ in terms of the eigenstates of the generator 
$\tilde{R}$. Then it might become possible to compute correlation functions 
using the thermofield theory. All these issues must be investigated further.  

\section{Conformal Scalar Field in 3D Schwarzschild-de-Sitter Space}
\hspace{5mm}

In this section the retarded Green function for a conformally coupled 
scalar field in 3 dimensional Schwarzschild-de-Sitter space (SdS$_3$) is 
considered by using the method of \cite{AHH} and the corresponding 
conformally invariant quantum mechanics model is derived. 

Schwarzschild-de-Sitter spacetime in d+1 dimensions (SdS$_{d+1}$) has the 
metric
\begin{eqnarray}
ds^2= -\tilde{f}(r) \, dt^2+\frac{1}{\tilde{f}(r)} \, dr^2+r^2 \, 
d\Omega_{d-1}^2,
\end{eqnarray}
where 
\begin{eqnarray}
\tilde{f}(r)= 1-2M \, r^{2-d}-\left(\frac{r}{\ell}\right)^2
\end{eqnarray}
and $M$ is a constant. When $d=2$ (and $d=1$), the rescaled metric $d\tilde{s}^2=ds^2/r^2$ 
coincides with that of $AdS_2 \times S_{d-1}$ and a hidden $SL(2, {\mathbb R})$ 
symmetry can be expected.\cite{AHH}
For $d=2$ the scalar field with angular momentum $l$ in this 
gravitational background obeys the equation of motion 
\begin{eqnarray}
&&\tilde{f}(r) \, \frac{d^2}{dr^2} \,  \phi(r) +\left(\frac{1-2M}{r} 
-\frac{3}{\ell^2} \, r\right) \, \frac{d}{dr}\phi(r) \nonumber \\
&& \qquad  +\left( -\frac{d^2}{4\ell^2}+\frac{x^2}{\ell^2}
-\frac{l^2}{r^2}+\frac{\omega^2}{\tilde{f}(r)} \right) \, \phi(r)=0.
\label{SdSd+1}
\end{eqnarray}
Here $\omega$ is the frequency and $x$ is a parameter defined in (\ref{x}).  
For the angular momentum $l$ is used instead of $m$ to avoid confusion with the 
mass. 

The angular momentum takes the value $l=0, \pm 1, \pm 2, ...$ 
For $l \neq 0$, there are two solutions, the normalizable and non-normalizable 
ones, which satisfy the ingoing boundary condition at the cosmological 
horizon. These are expressed in terms of the hypergeometric functions as 
follows. 
\begin{eqnarray}
\phi_{n.}(r) &=& ( 1-z)^{-i\frac{\ell^2 \omega}{2a}} \, z^{\frac{|l|\ell}{2a}}
\, F\left(\alpha_+,\alpha_-;\gamma ;z\right), 
\nonumber \\
\phi_{n.n.}(r) &=& ( 1-z)^{-i\frac{\ell^2 \omega}{2a}} \, 
z^{-\frac{|l|\ell}{2a}}
\, F\left(\alpha_+-\gamma+1,\alpha_--\gamma+1; 2-\gamma;z\right)  \label{solF}
\end{eqnarray}
Here  $z=(r/a)^2$ and 
\begin{eqnarray}
a=\ell \, \sqrt{1-2M}, \qquad \alpha_{\pm}= \frac{1 \pm x}{2}+\frac{|l|\ell}{2a}
-i\frac{\ell^2\omega}{2a}, \qquad 
\gamma=1+\frac{|l|\ell}{a}
\end{eqnarray}
In SdS$_3$ the Hawking temperature is given by $T_{SdS}=a/(2\pi \ell^2)$.\cite{GH} 
Then the retarded Green function can be calculated as in \cite{AHH}. 
For the case of the conformal coupling $x=1/2$ it turns out to be given by
\begin{eqnarray}
G^R_l(t) \propto \theta (t) \, \left( \frac{1}{\sinh \frac{at}{2 \, \ell^2}} 
\right)^{1+\frac{2 |l|\ell}{a}}. \label{GRSdS}
\end{eqnarray}
To derive this result for the mode $l \, $ ($\neq 0$), one requires that 
$\phi_l=B \, \phi_{n.}+A \, \phi_{n.n.}$ be ingoing at the horizon $z=1$.
This determines the ratio $B/A$ and one obtains 
\begin{eqnarray}
\tilde{G}^R_l(\omega) \propto \frac{B}{A}=-\frac{\Gamma(1-\frac{|l|}{a}\ell)}
{\Gamma(1+\frac{|l|}{a}\ell)} \, \frac{\Gamma(\frac{1}{2}
+\frac{|l|}{a}\ell-i \frac{\ell^2}{a}\omega)}{\Gamma(\frac{1}{2}
-\frac{|l|}{a}\ell-i \frac{\ell^2}{a}\omega)}. 
\end{eqnarray}
With a non-vanishing $M$ the residues at the poles are regularized. 

For the mode $l=0$, the solutions (\ref{solF}) in terms of hypergeometric 
functions are degenerate and another solution which includes $\log r$ must be 
taken into account. This case needs some care. 
For this purpose one rewrites $\phi_l$ as
\begin{eqnarray}
\phi_l &=& D \, ( 1-z)^{-i\frac{\ell^2 \omega}{2a}} \, z^{\frac{|l|\ell}{2a}}
\, F\left(\alpha_+,\alpha_-;\gamma ;z\right) \nonumber \\
&&+ C \, \frac{1}{1-\gamma} \, (1-z)^{-i\frac{\ell^2 \omega}{2a}} \, \left[
z^{-\frac{|l|\ell}{2a}}
\, F\left(\alpha_+-\gamma+1,\alpha_--\gamma+1; 2-\gamma;z\right) \right. 
\nonumber \\
&& \qquad \qquad \left. - z^{\frac{|l|\ell}{2a}}
\, F\left(\alpha_+,\alpha_-;\gamma ;z\right)\right] \label{l0}
\end{eqnarray}
Here $C= (1-\gamma) \, A$ and $D=A+B$. 
One then analytically continues the 
variable $l$ from an integer variable to a continuous one, and takes the limit 
$l \rightarrow 0$. Then $\gamma \rightarrow 1$ and one obtains
\begin{eqnarray}
\tilde{G}^R_0(\omega) \propto \lim_{l \rightarrow 0} \, \frac{D}{C}= 2 \, 
\psi \left(\frac{1}{2}
-i\frac{\ell^2}{a}\omega \right)-\psi(1)-2 \, \log 2,  
\end{eqnarray}
where $\psi(z)$ is a di-gamma function. By discarding the analytic terms, 
which gives contact terms, and performing Fourier transform, one gets 
(\ref{GRSdS}) for $l=0$. Note that not only the first term but also the second 
in (\ref{l0}) is actually normalizable for $l \rightarrow 0$, but the second
term represents the source and the first term the response.

From the above result one can read off the conformal weight. 
\begin{eqnarray}
\tilde{\Delta}= |l| \, \frac{\ell}{a}+\frac{1}{2}= \frac{|l|}{\sqrt{1-2M}}
+\frac{1}{2} 
\end{eqnarray}
Compared to (\ref{sdimension}) with $d=2$ substituted, the conformal weight is 
smoothly deformed by $M$. 
The conformal quantum mechanics model which reproduces this Green function can 
be derived by the same procedure as in sec 4. In the present case 
$r_0=\tilde{\Delta}$ and 
\begin{eqnarray}
(\tilde{\Delta}-\frac{1}{2})^2= \frac{1}{4} \, \frac{l^2}{1-g_0}+\frac{1}{2} 
\, g_1
\end{eqnarray}
This is satisfied, if
\begin{eqnarray}
g_0=\frac{3}{4} +\frac{1}{2}\,M, \qquad g_1=0. 
\end{eqnarray}
The corresponding conformal quantum mechanics theory is defined by the 
Lagrangian 
\begin{eqnarray}
L_{SdS_3}= \frac{1}{8} \, \dot{\bm{x}}^2-\left(\frac{3}{8}
-\frac{M}{4} \right) \, \frac{(\bm{x} \cdot \dot{\bm{x}})^2}{\bm{x}^2}. 
\end{eqnarray}

In the above example a deformation of the background metric of spacetime 
leads to a deformation of the parameter $g_0$ of the conformal quantum 
mechanics. It can be said that this Lagrangian depends on the Hawking 
temperature $T_{SdS}$ via the parameter $M$. For higher dimensions 
($d >2$) there will be no $SL(2,{ \mathbb R})$ symmetry. It would be 
interesting if a non-conformal quantum mechanics 
model which corresponds to a scalar field in $SdS_{d+1}$ could be found.

\section{Large-N Matrix Model}
\hspace{5mm}
In this section the results of sec.4  are extended to 
the large N matrix model. It is assumed that the static patch observer 
is described by N by N hermitian matrices $X^i$ ($i=1,..,d$). 
It may be a gauged D=1 matrix model. In this case the gauge field matrix 
$A_{\tau}$ can be gauged fixed, $A_{\tau}=0$, and its only role will be to 
impose a vanishing charge condition $Q=0$ on the state vectors. In the 
diagonal matrix approximation which will be adopted here, this condition is 
nothing but the permutation symmetry of the eigenvalues of the matrices. 
Hence a gauge field will not be considered here. 

The following form of the Lagrangian is assumed. 
\begin{eqnarray}
L_M= \frac{1}{2}\,  \mbox{Tr} \sum_{i=1}^d \, (\dot{X}^i)^2
-\frac{g_2 }{\mbox{Tr} \, \sum_{i=1}^d \, (X^i)^2}
\end{eqnarray}
Here $g_2$ is a constant. In what follows the diagonal approximation will be 
adopted, {\em i.e.}, the matrix $X^i$ is assumed to be diagonal. In the 
classical geometric limit, the off-diagonal matrix elements are assumed to be 
heavy compared to the diagonal elements. 
The diagonal elements are denoted as $x_a^i(\tau)$, ($a=1,2,...,N$).  Then
the Lagrangian reduces to  
\begin{eqnarray}
L_M=\frac{1}{2} \, \sum_a \, \dot{\bm{x}}_a^2-\frac{g_2 }{\sum_a \bm{x}_a^2}.
\label{MatrixLag}
\end{eqnarray}
The momentum conjugate to $x_a^i$ is given by $p_a^i=\dot{x}_a^i$. The
 conserved charges at $\tau=0$ are 
\begin{eqnarray}
H &=& \frac{1}{2} \, \sum_a \bm{p}_a^2+\frac{g_2 }{\sum_a \bm{x}_a^2}, 
\label{Hmatrix} \\
D &=&  -\frac{1}{2} \, \sum_a \, \bm{x}_a \cdot \bm{p}_a, \\
K &=& \frac{1}{2} \, \sum_a \, \bm{x}_a^2
\end{eqnarray}

Now it is assumed that the primary state has the wave function, 
\begin{eqnarray}
\psi(\bm{x}_a)= \left(\sum_a \, c_{i_1,i_2, ..., i_l} \, x_a^{i_1} 
\cdots x_a^{i_l} \right)\ \left(\sum_b \bm{x}_b^2\right)^{\alpha} \ 
\exp \left
\{\frac{-1}{2} \
\sum_c \bm{x}_c^2\right\}
\label{MatrixPrimary}
\end{eqnarray}
where $c_{i_1,..,i_l}$ is a constant symmetric traceless tensor,  
and $\alpha$ is a constant.  
The lowest-weight condition $L_- \, \psi=0$ yields the equation.
\begin{eqnarray}
\alpha=r_0-\frac{l}{2}-\frac{1}{4} \, Nd, 
\label{alpha}
\end{eqnarray}
where $r_0$ is the eigenvalue of $R$. 
Now the eigenvalue equation $R \psi=r_0 \psi$ gives
\begin{eqnarray}
R \psi= \frac{1}{4} \, (2l+4\alpha+Nd)\psi+\frac{1}{2\sum_a \bm{x}_a^2} \, 
(g_2-2l\alpha-Nd\alpha-2\alpha(\alpha-1)   )\psi.
\end{eqnarray}
From this one obtains two constraints.
\begin{eqnarray}
r_0 &=& \frac{1}{4} \, (2l+Nd+4\alpha), \\
g_2  &=& (2r_0+l+\frac{Nd}{2}-2)(r_0-\frac{l}{2}-\frac{1}{4}Nd)
\end{eqnarray}
The first equation is consistent with (\ref{alpha}).
To obtain the eigenvalue $r_0=l+\frac{d-1}{2}$ (\ref{sdimension}), 
one must set 
\begin{eqnarray}
\alpha &=& \frac{l}{2}+\frac{d-1}{2}-\frac{N}{4} \, d, \\
g_2 &=& \frac{3}{2} \, l(l+d-2)-\frac{d}{2} \, (N-1) \, l-
\frac{1}{8} \, [Nd+2(d-3)][Nd-2(d-1)]
\end{eqnarray}
Note that this value of $\alpha$ ensures the normalizability of $\psi$. 
One must note that the constant $g_2$ should not depend on the angular 
momentum $l$. So the combination $l(l+d-2)$ is to be relpaced by the 
squared angular momentum operator, 
\begin{eqnarray}
\bm{L}^2=\sum_{a,b} \, 
[(\bm{x}_a \cdot \bm{x}_b) \, (\bm{p}_a \cdot \bm{p}_b)
-(\bm{x}_a \cdot \bm{p}_a)(\bm{x}_b \cdot \bm{p}_a)]. 
\end{eqnarray}
The operator 
ordering must be correctly specified. There is also a linear term of $l$. 
This must be also replaced by 
\begin{eqnarray}
l \rightarrow -\frac{d-2}{2}+\sqrt{\frac{1}{4}(d-2)^2+\bm{L}^2 }.
\end{eqnarray}
Finally, the hamiltonian $H$ (\ref{Hmatrix}) is given by the following 
complicated expression. 
\begin{eqnarray}
H_M &=& \frac{1}{2} \, \sum_a \, \bm{p}_a^2+\frac{3}{2\sum_c \, \bm{x}_c^2} 
\,   \bm{L}^2  -\frac{d(N-1)}{2\sum_b\bm{x}_b^2} \, \sqrt{\frac{(d-2)^2}{4}
+\bm{L}^2  }
\nonumber \\
&& + \frac{1}{\sum_b \, \bm{x}_b^2}\left [\frac{d(d-2)}{4}(N-1)-\frac{1}{8}
 \, (Nd+2d-6)(Nd-2d+2)\right].
\label{MatrixH}
\end{eqnarray}
The corresponding Lagrangian $L'_M$  then differs from (\ref{MatrixLag}). 
This can be obtained by eliminating $p_a^i$ from $L'_M= \sum_a \dot{\bm{x}}_a 
\cdot \bm{p}_a-H_M$. This Lagrangian is implicitly given by
\begin{eqnarray}
L'_M &=& \frac{1}{2} \, \sum_a \bm{p}_a^2 
+\frac{1}{\sum_c \, \bm{x}_c^2} \, \left[\frac{3}{2} \,  \bm{L}^2
+\frac{N-1}{8}\frac{d(d-2)^2}{\sqrt{\frac{1}{4} \, (d-2)^2
+ \bm{L}^2}} \right.\nonumber \\
&&\left.-\frac{1}{4} \, d(d-2)(N-1)+\frac{1}{8} \, \{Nd+2d-6)(Nd-2d+2)\}  
 \right],
\end{eqnarray}
where $p_a^i$ is determined by solving the following equation for $p_a^i$
\begin{eqnarray}
\dot{x}_a^i= p_a^i+\frac{1}{\sum_c \, \bm{x}_c^2} \, \left[3-\frac{d(N-1)}{2} \,
\frac{1}{\sqrt{\frac{1}{4} \, (d-2)^2+ \bm{L}^2}}\right] \, \sum_b 
\, [(\bm{x}_a \cdot \bm{x}_b)p_b^i-(\bm{x}_a \cdot \bm{p}_b)x_b^i].
\end{eqnarray}
It is  straightforward to obtain the matrix form of the Hamiltonian from 
(\ref{MatrixH}). 

\section{Discussion}
\hspace{5mm}
In this paper a simple conformal quantum mechanics model of d-component 
fields, $x_i(\tau)$, ($i=1,..,d$) (\ref{Lagrangian1}) is proposed, which 
exactly reproduces the retarded Green functions and conformal weights 
of scalar fields with conformal coupling and four dimensional gravitons seen 
by a static patch observer. The model obtained in sec.4 is not of the type of 
large N matrix model. 

In sec.5 the Lagrangian of the quantum mechanics model obtained in the 
previous section is rewritten in terms of the de Sitter time $t$. It is 
found that the action integral is given by a contour integral along a closed 
contour in the complex-$t$ plane. This is nothing but the Schwinger-Keldysh
formalism of thermofield theory. Hence the finite temperature is naturally 
encoded in the conformal quantum mechanics model. 

The conformal quantum mechanics model we found is extended to a large 
N matrix model in sec.7. This model, however, has several problems. The 
hamiltonian is obtained, but the Lagrangian is complicated and its form is 
worked out only implicitly. The form of the matrix model Hamiltonian is 
unusual one with the traces of matrices in the denominators. Although the 
primary states of the form (\ref{MatrixPrimary}) are uniquely determined and 
will correspond to the conformally coupled scalar fields in the static patch, 
there may be more primary states. It is important to identify the whole 
primary states. In this respect an analysis of the full matrix model is 
important. On the other hand, the construction of the large N matrix model 
in sec.7 can be applied to the $x \neq 1/2$ case with $SL(2,{\mathbb R}) 
\times SL(2,{\mathbb R})$ symmetry. 

In sec.6 a conformal quantum mechanics model for a scalar field with a 
conformal coupling in SdS$_3$ is constructed and it is found that the 
Lagrangian depends on the Hawking temperature $T_{SdS}$ via the parameter $M$. 
A similar analysis can be applied to the $x \neq 1/2$ case with 
$SL(2,{\mathbb R}) \times SL(2,{\mathbb R})$ symmetry\cite{AHH} and  
will lead to an interacting Lagrangian which depends on $T_{SdS}$ or $M$.

\end{document}